\documentclass{article}
\usepackage{spconf,amsmath,epsfig,graphicx,listings,hyperref}

\title{TOWARDS EXASCALE REAL-TIME RFI MITIGATION}
\name{Rob V. van Nieuwpoort}
\address{Netherlands eScience center and University of Amsterdam, The Netherlands\\
        R.vanNieuwpoort@esciencecenter.nl, R.vanNieuwpoort@uva.nl}
\begin{document}
%
\maketitle
\begin{abstract}
We describe the design and implementation of an extremely scalable
real-time RFI mitigation method, based on the offline AOFlagger. All
algorithms scale linearly in the number of samples.  We describe how
we implemented the flagger in the LOFAR real-time pipeline, on both
CPUs and GPUs. Additionally, we introduce a novel simple history-based
flagger that helps reduce the impact of our small window on the data.

By examining an observation of a known pulsar, we demonstrate that our
flagger can achieve much higher quality than a simple thresholder,
even when running in real time, on a distributed system. The flagger
works on visibility data, but also on raw voltages, and beam formed
data.  The algorithms are scale-invariant, and work on microsecond to
second time scales.  We are currently implementing a
prototype for the time domain pipeline of the SKA central signal
processor.
\end{abstract}
\section{Introduction}
\label{sec:intro}

Radio Frequency Interference (RFI) mitigation is extremely important
to take advantage of the vastly improved bandwidth, sensitivity, and
field-of-view of exascale telescopes. For current instruments, RFI
mitigation is typically done offline, and in some cases (partially)
manually.  At the same time, it is clear that due to the high
bandwidth requirements, RFI mitigation will have to be done
automatically, and in real-time, for exascale instruments.

In general, real-time RFI mitigation will be less precise than offline
approaches. Due to memory constraints, there is much less data to work
with, typically only in the order of one second or less, as opposed to
the entire observation. In addition, due to memory limitations and the
fact that processing is typically done in a distributed system, we can
record only limited statistics of the past. Moreover, we will
typically have only few frequency channels locally available at each
compute node. Finally, the amount of processing that can be spent on
RFI mitigation is extremely limited due to computing constraints and a
limited power budget. Many existing algorithms are therefore far too
expensive and not applicable.

Nevertheless, there are many potential benefits as well, which include
the possibility of working on higher time and frequency resolutions,
before any integration is done, leading to more accurate results. Most
importantly, we can remove RFI before beam forming, which combines
data from all receivers.  With beam forming, not only the signals, but
also the RFI that is present in the data streams from the separate
receivers is combined, effectively taking the union of all RFI. Thus,
the RFI from any receiver will pollute all beams. Therefore, it is
essential to also perform real-time RFI mitigation before the beam
former, even though the data rates can be very high at this point.
This is particularly important for pulsar surveys, for instance.

The algorithms we use are based on earlier work by Offringa and others~\cite{andre-post,vandeGronde2016,lofar-results}.
Although our techniques are generic, we describe how we
implemented real-time RFI mitigation for one of the SKA pathfinders:
The Low Frequency Array (LOFAR)~\cite{lofar-performance}. The modified RFI mitigation
algorithms we introduce here are extremely fast, and the computational
requirements scale linearly in the number of samples and frequency
channels. We evaluate the quality of the algorithms with LOFAR pulsar
observations.  Using the signal-to-noise ratios of the folded pulse
profiles, we can qualitatively and quantitatively compare the impact
of different real-time RFI mitigation algorithms.

In addition to CPU versions, we have developed a prototype
for Graphical Processing Units (GPUs). We present the very
promising performance results of performing real-time RFI mitigation
on GPUs. Finally, we are now working on incorporating our
CPU and GPU codes in the SKA Central Signal Processor (CSP), in the
context of the Time Domain Team (TDT) pulsar and transient pipeline.

\section{LOF: THE LOFAR ONLINE FLAGGER}

For the offline RFI mitigation for the LOFAR telescope, we use the
AOFlagger~\cite{lofar-results}. This flagger is relatively efficient, and is fast
enough to be applied in modern high-resolution observatories. The
AOFlagger operates on visibility data, and currently is used only in
the imaging mode.  Although the AOFlagger provides a selection of
many different algorithms, for online use, we ported the most important
and successful ones to the LOFAR real-time central processing system,
that originally ran on an IBM Blue Gene/p supercomputer. We also
ported the algorithms to NVIDIA GPUs using CUDA, as described
in Section~\ref{sec:gpu}. We call our implementation of the real-time flagger
“LOF”, short for \emph{LOFAR Online Flagger}.

\subsection{The SumThreshold algorithm}

\begin{figure}
\centering
\begin{minipage}{.6\columnwidth}
\centering
\includegraphics[width=.95\columnwidth]{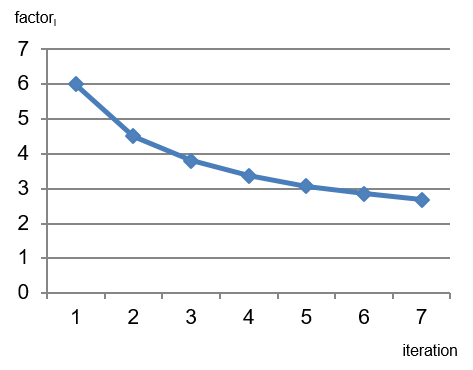}
\end{minipage}%
\begin{minipage}{.4\columnwidth}
\centering
\includegraphics[width=.95\columnwidth]{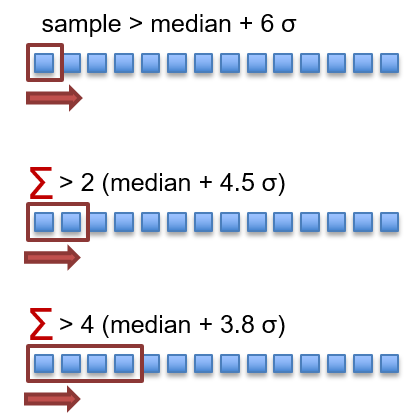}
\end{minipage}
\caption{Left graphs shows the default $factor_I$ for different iterations; right graph shows the operation of SumThreshold. The sliding window doubles every iteration.}
\label{fig:SumThreshold}
\end{figure}

The most important algorithm we use is SumThreshold~\cite{andre-post}. It performs
thresholding with an exponentially increasing window size, and an
increasingly sharper threshold. This way, it can detect RFI at
different scales. As we will demonstrate in Section~\ref{sec:eval},
this indeed works well in practice, and effectively removes RFI
from microsecond to multiple second scales.  With SumThreshold, we
define the threshold for the current window as follows: \\
$threshold_I = median + stddev * factor_I * sensitivity$ \\
Where the factor is: $factor_I = \frac{startThreshold * p^2log(I)}{I}$ \\
  \\
Typical values are p = 1.5, and sensitivity = 1.0. All measurements 
use the defaults, since we empirically found that they provide optimal results. No tuning is required. Figure~\ref{fig:SumThreshold} shows the threshold for different iterations.

Figure~\ref{fig:SumThreshold} shows the operation of the SumThreshold algorithm. With the
default values, the algorithm begins the first iteration as a simple
thresholder, flagging all single samples that are more than 6~sigma away
from the median. The second iteration doubles the window size to two
samples, but lowers the threshold to 4.5~sigma. We typically run 7-10~iterations,
depending on the resolution and size of the input.

SumThreshold can be run in a one-dimensional mode, as
shown in Figure~\ref{fig:SumThreshold}. This can then be done in both time and frequency
directions. Alternatively, the algorithm can operate on 2D data in one
pass as well (not shown in Figure~\ref{fig:SumThreshold}). As we will explain in Section~\ref{sec:rt-changes},
for real-time use, we did implement a 2D code, but we will mostly
use the one dimensional version of the algorithm for performance
reasons.  It is important to note that the computational complexity of
the SumThreshold algorithm itself is linear in the number of
samples. Therefore, the algorithm is suitable for real-time use, where
we are limited by the number of compute cycles we can spend.

\subsection{The scale-invariant rank operator}

\begin{figure}[htb]
\center
\includegraphics[width=0.8\columnwidth]{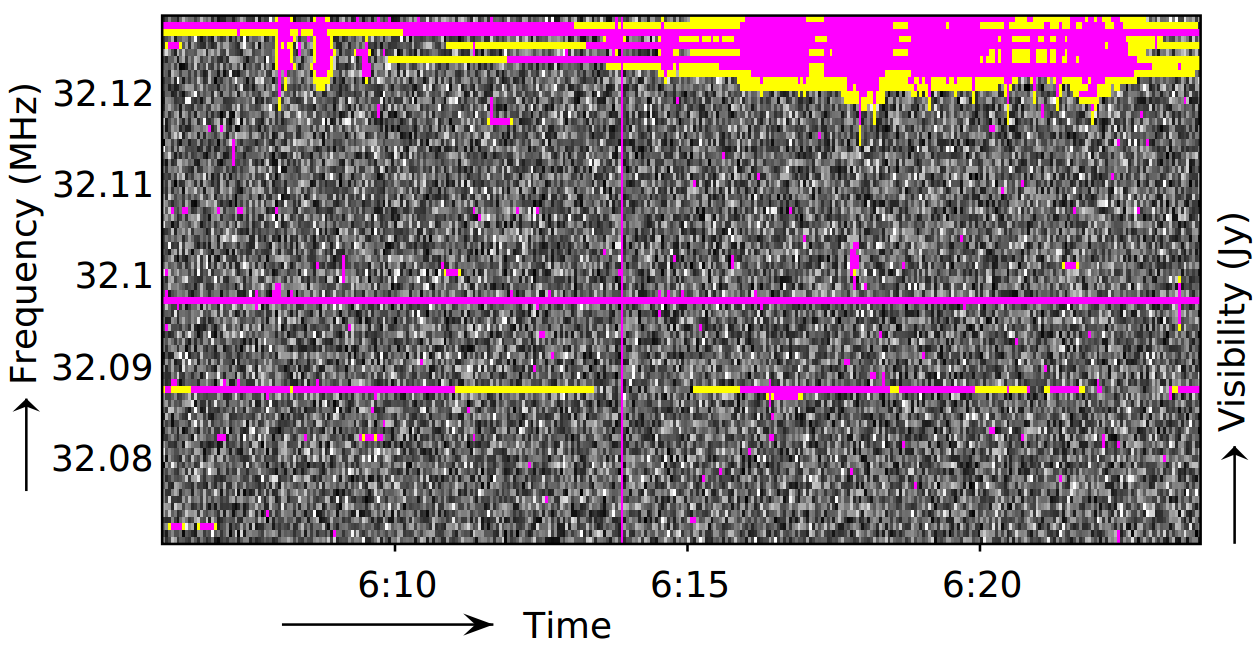}
\caption{Result of running the SIR operator. Purple samples are flagged by SumThreshold; yellow samples are subsequently flagged by the SIR operator. Image courtesy Andre Offringa.}
\label{fig:SIR}
\end{figure}

The scale-invariant rank operator~\cite{vandeGronde2016} makes
SumThreshold more robust, by extending the ranges of flagged samples
by a percentage of the size of the flagged range. A typical percentage
we used is 20\%. This means that all ranges of consecutive flagged
samples are extended by flagging 20\% more samples, before and after
the original range. The algorithm can be run in both the time
direction and the frequency direction.  The SIR operator helps to
remove RFI that slowly rises and decreases in strength, that may be
otherwise undetected. The result of the SIR operator is shown in
Figure~\ref{fig:SIR}.

We use a version of the SIR operator implementation that has a
linear computational complexity in the number of samples (the original
implementation in the AOFlagger had worse computational
complexity). The linear version of the algorithm is described in~\cite{vandeGronde2016}.
Since SIR operates on the flag masks only, and not on the actual
data itself, it is extremely efficient.

\section{INTEGRATION IN REAL-TIME PIPELINES}

\begin{figure}[htb]
\includegraphics[width=\columnwidth]{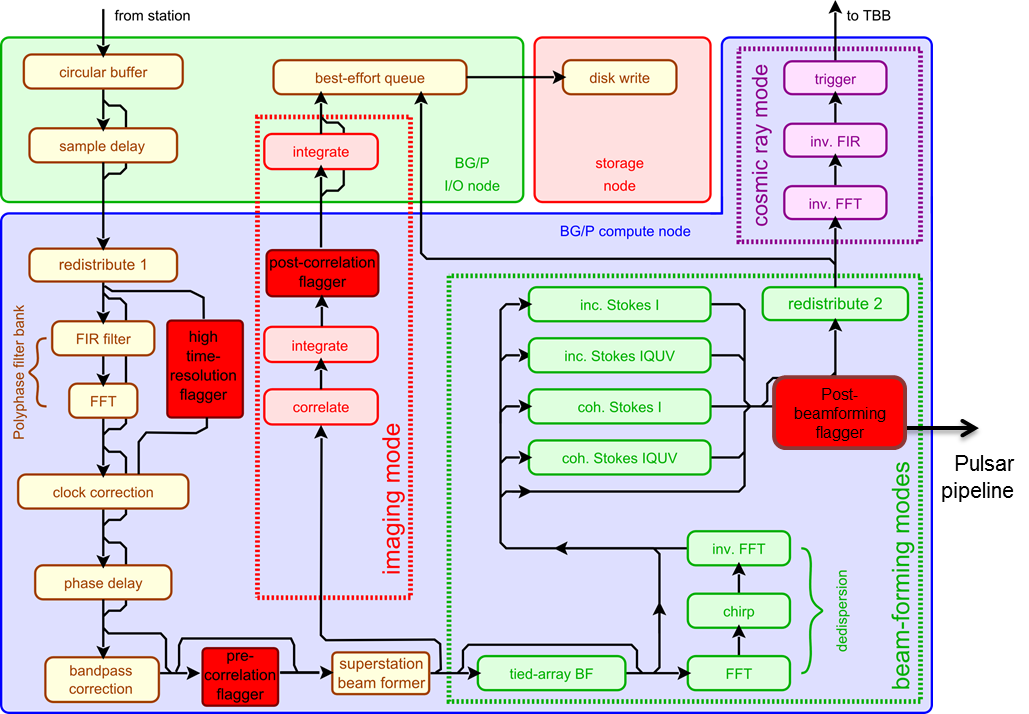}
\caption{The LOFAR real-time pipelines. Input data rates vary between 400 gbit/s and 1.6 tbit/s for the pre-correlation flagger, depending on the mode (4, 8 or 16 bit samples).}
\label{fig:LOFAR}
\end{figure}

Figure~\ref{fig:LOFAR} shows a high-level overview of the LOFAR
real-time central signal processing pipeline. The entire pipeline is
implemented in software, and is described in detail in~\cite{lofar-performance}, including
a detailed performance analysis. The LOFAR online flagger components
are placed in four different places in the pipeline. Depending on the
configuration and the observation type, one or more different flaggers are used.

Data arrives in the form of raw voltages at the top left of the
figure. Next, a number of steps are executed independently of whether
we are in imaging or beam forming mode. The most important step
is a polyphase filter bank that splits the broad input subbands in
narrower frequency channels. Typical channel bandwidths are between
0.8~kHz and 12~kHz, with sample rates between 82~microseconds
and 1.3~milliseconds. When we need extremely high time
resolution (e.g, for millisecond pulsars or for the cosmic ray
pipeline), we bypass the polyphase filter bank altogether. For this
case we created a special high time resolution flagger.

Next, the band pass of the first polyphase filter bank that runs
inside the stations on FPGAs is corrected. After this bandpass
correction, we inserted our pre-correlation flagger that works on the
channelized raw voltage data. It is important to do this after the
band pass correction, as this ensures that the sensitivity is equal
across all channels. The pre-correlation version is the most important
real-time flagger, especially for the beam forming modes. The beam
former does a weighted addition of the data streams from the different
stations, essentially taking the union of all RFI from all
stations. If RFI is present at a station, this will pollute all
output beams. Especially for uncorrelated RFI and long baselines, this is sub optimal.

In addition, we can use our real-time flagger after the correlator, for
real-time image-based transient detection, for example.
The drawback of performing RFI mitigation after the
correlator is that, depending on the number of baselines and the
integration time, data rates can be higher than before the
correlator. Finally, we have a post beam forming flagger that can
potentially benefit from a better interference-to-noise
ratio (INR). Moreover, depending on the number of output beams, this
typically runs on lower data rates.

\section{CHANGES FOR REAL-TIME USE}
\label{sec:rt-changes}

To make sure the algorithms used in the AOFlagger work in a real-time
context, we had to make several changes. First, depending on the input
data type of the flagger, we may have to compute amplitudes first. This is
the case for the high-time resolution flagger, and for the
pre-correlation flagger. The post-correlation flagger runs on
visibility data directly, while the post-beam forming flagger runs on
Stokes~I data.

\begin{figure}[htb]
\center
\includegraphics[width=.5\columnwidth]{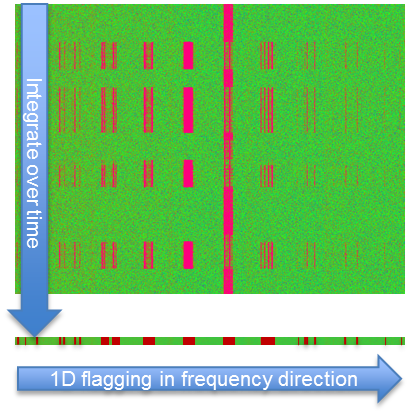}
\caption{Integration in the time direction to perform flagging in the frequency direction with high INR.}
\label{fig:integrate}
\end{figure}

In pre-correlation mode, we typically integrate the data. We found
it particularly useful to integrate the time direction
fully for frequency flagging, and to integrate the frequency direction
fully for time domain flagging. Figure~\ref{fig:integrate} shows this. This
approach has two benefits: it improves the INR, while at the same time reducing the computational costs. 
We first flag in the frequency direction. This removes strong narrow-band
RFI that would otherwise decrease the quality of the statistics used
to compute the thresholds. We found that this is frequently present in
the LOFAR RFI environment. This also is the reason we create the narrow approximately 1kHz channels.
We have an alternative method that implements
2D~flagging, while partially integrating in one or both
dimensions to improve INR and to reduce compute costs.

All flaggers have linear computational complexity in
the number of samples, regardless of their place in the real-time
pipeline. For instance, the pre-correlation flagger has a complexity
of \emph{O(nrStations * nrPolarizations * nrChannels * nrTimes)}, the post-correlation
flagger is \emph{O(nrBaselines * nrPolarizations * nrChannels)}, and the post beam
forming flagger is \emph{O(nrBeams * nrChannels * nrTimes)}.

\subsection{Using historical information}

One of the most difficult problems with real-time RFI mitigation is
the very limited window on the observation. Typically, we can only
keep one second of data or less (e.g., a tenth of a second) in memory.
Similarly, we can only keep a very limited number of frequency
channels in memory. This is due to memory constraints, and
partially because we process the data on a distributed
system. We create parallelism by performing domain
decomposition. This typically means that different frequency subbands
are processed on different compute nodes.
Finally, in some cases, processing a second of input data
takes longer than a second. To still meet the real-time requirements,
subsequent seconds are processed partially overlapping in time,
by different compute nodes. All these factors severely limit our
situational awareness.

Our solution for this problem is the introduction of a novel history
flagger that performs simple thresholding of the current data chunk,
based on statistics of past chunks. Pseudo code for this history
flagger is shown below.
 \\
\begin{lstlisting}
// For all channels, we do the following:
// Keep a history buffer (sliding window) of
// means of unflagged samples of the past seconds

currentValue = meanOfUnflaggedSamples()

historyMean = meanOfMeans()
historyStddev = stddevOfMeans()
threshold= historyMean + sensitivity * historyStddev

if(currentValue < threshold) {
    addToHistory(station, subband, currentValue)
} else {
    addToHistory(station, subband, threshold)
    flagThisIntegrationTime()
}
\end{lstlisting}
\vspace{-1cm}

\begin{figure}[tb]
\center
\includegraphics[width=.7\columnwidth]{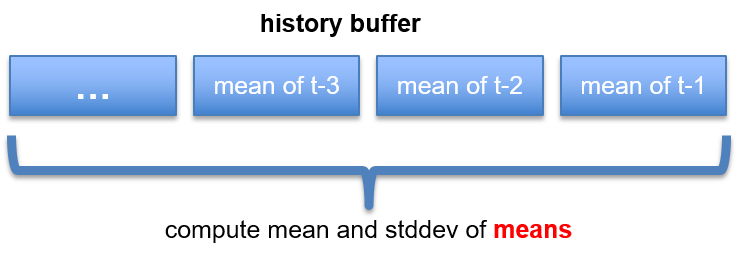}
\caption{The history buffer.}
\label{fig:history}
\end{figure}

As shown in Figure~\ref{fig:history}, our flagger uses a history
buffer that stores the means of the unflagged samples of the previous
seconds of data. This buffer essentially is a sliding window over the
data. We use the buffer to compute the mean and standard deviation of
the previous seconds, to give us a frame of reference for the overall
signal strength of the current second. This is especially important,
since a strong broadband RFI event that lasts longer than our
integration time can otherwise not be detected.

\subsection{Statistics in a real-time pipeline}
The quality of the statistics that we compute and keep is important,
especially since the window on the data is so small. We use only very
basic statistics, such as (winsorized) means, medians, standard
deviations, and the MAD (Median Absolute Deviation). To compute the
medians, which can be expensive, we use an efficient \emph{O(N)}
implementation (note that sorting the data already is \emph{O(NlogN)} at
best. More complications result from using a distributed 
platform: statistics are often in the wrong place,
at the wrong time. Moreover, we cannot compute running statistics,
since a second of data takes more than a second to compute. Finally,
our real-time pipeline uses complex communication patterns due to
scheduling, and asynchronous communication for better performance.
Together, this means that even
computing basic statistics is quite complex in practice.

An important consideration for the history flagger is presented by the
space requirements of the statistics of the past we want to keep in
the history buffer. Let us use LOFAR numbers as an example. For the
pre-correlation flagger, we need $stations * subbands * channels *
32 bits = 64 * 248 * 256 * 4$ = 15.5~MByte per second. If we want
to keep 5 minutes of history in the buffer, we already need 300~samples,
leading to a storage requirement of 4.5~GBytes. After the
correlator, requirements are even higher: $baselines * subbands *
channels * 32 bits = 2080 * 248 * 256 * 4$ = 504~MByte per second,
which, even if we want to store only 5 minutes, already leads to 148~Gbytes
of statistics data. Therefore, in practice, even keeping these very
limited statistics of only a few minutes of the past already is
extremely difficult, and down-sampling may be needed.

\section{evaluation}
\label{sec:eval}

In this section, we will present a qualitative and quantitative
evaluation of the real-time flagger, using a LOFAR pulsar
observation.
We use the pulsar
pipeline, because it allows for a quantitative comparison: we perform
dedispersion and folding to create a folded pulse profile. Next, we
compute the SNR of the pulse profile as a measure of quality. Better
RFI mitigation directly leads to a higher SNR.

\subsection{A pulsar observation}

We performed an observation of pulsar B1919+21, which has a period of
1.3373~s, a pulse width 0.04~s, and a dispersion measure (DM) of
12.455. We observed at 138.0–145.2 MHz (32 subbands) with 5~stations:
CS005, CS006, RS205, RS406, and UK608. We deliberately chose a
number of core stations, where RFI is expected to be correlated, a
remote station in the Netherlands, where the RFI environment is known
to be particularly bad, and an international station in the UK, to
guarantee we also have uncorrelated RFI.

We used a special LOFAR mode that allowed us to store the raw UDP
network packets, before the data even enters the correlator. This
allows us to replay the entire real-time pipeline in an offline mode,
enabling comparisons between different flagging algorithms, parameter
settings and even observation modes (e.g., imaging or beam forming). For the
beam forming mode that we use for the pulsar pipeline, we split the
frequency subbands into 16~channels (12~KHz / 82~μs).

\begin{figure}[t]
\center
\includegraphics[width=\columnwidth]{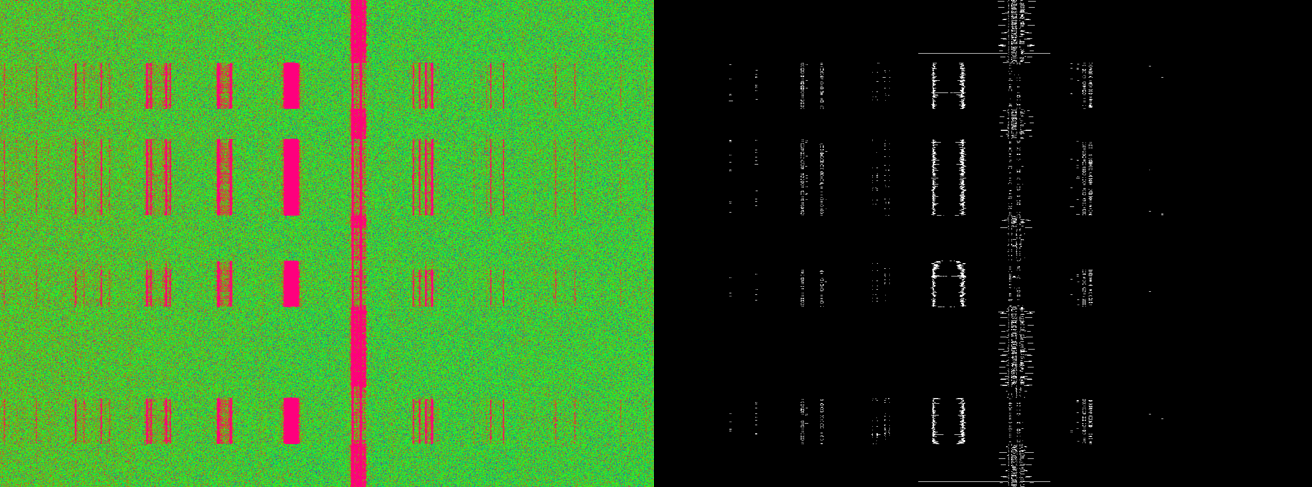}
\caption{Thresholding vs LOF, on raw voltages. Left: original data; right: difference threshold – LOF, i.e. RFI residuals that are flagged by LOF, but not by simple thresholding.}
\label{fig:waterfall}
\end{figure}

Figure~\ref{fig:waterfall} shows two waterfall plots. The left side is the original observation
with RFI present; on the right the improvement of our LOF compared to
a simple thresholding scheme, where we manually determined the optimal
threshold. There clearly is a lot of residual RFI that is not removed
by the thresholder, that is removed by LOF.

\begin{figure}[t]
\begin{minipage}{\columnwidth}
\centering
\includegraphics[width=0.8\columnwidth]{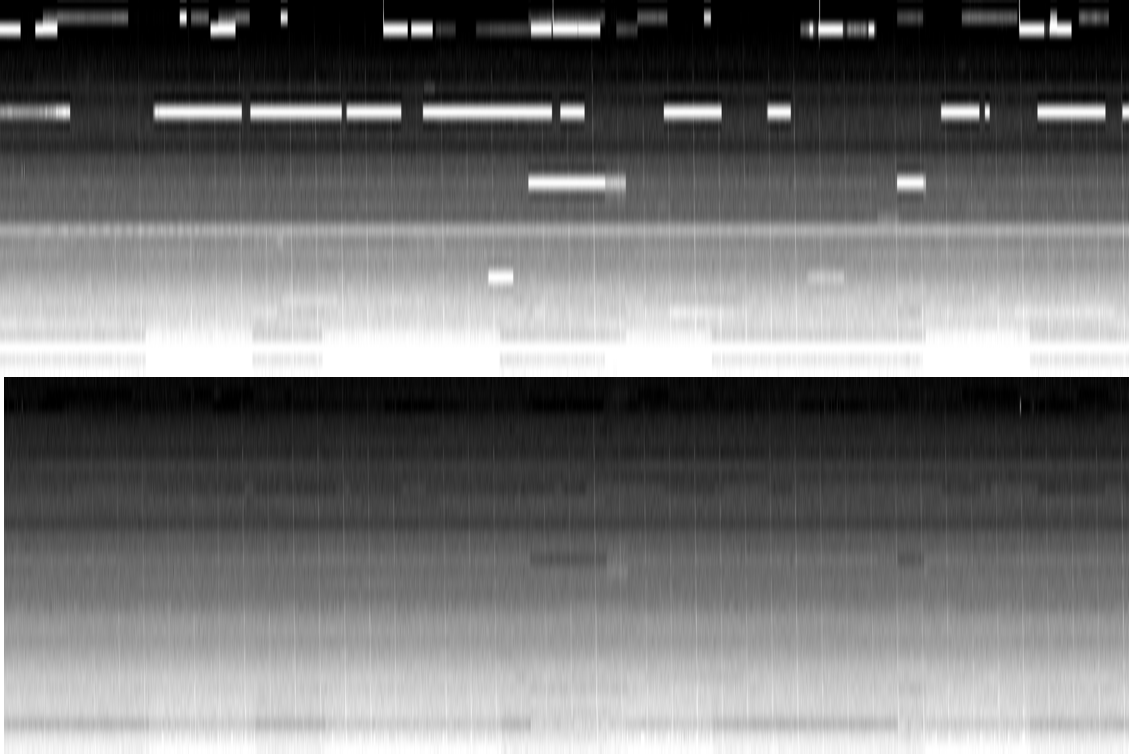}
\caption{Beam formed data (Stokes I), top panel without RFI mitigation, bottom panel with LOF.}
\label{fig:stokesI1}
\end{minipage}
\begin{minipage}{\columnwidth}
\centering
\includegraphics[width=0.8\columnwidth]{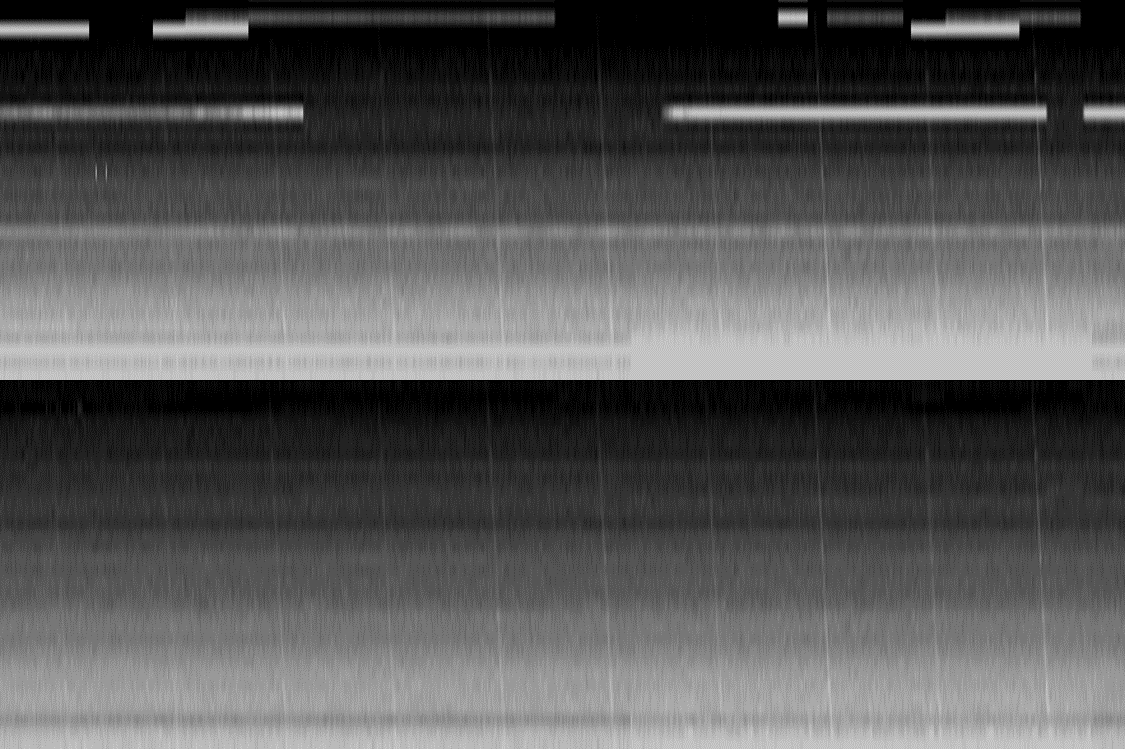}
\caption{Closeup of beam formed data (Stokes I), top panel without RFI mitigation, bottom panel with LOF.}
\label{fig:stokesI2}
\end{minipage}
\end{figure}



Figure~\ref{fig:stokesI1} shows Stokes~I data of an output beam; Figure~\ref{fig:stokesI2}
shows the same, but zoomed in. The top panels are without
RFI mitigation, the bottom panels with LOF. Almost all RFI is
removed. The data is not de-dispersed, but the pulsar signal is so
strong that it is clearly visible in the data. \emph{Note that the
pulses are not flagged away by our mitigation algorithms.}

\begin{figure}[t]
\center
\includegraphics[width=\columnwidth]{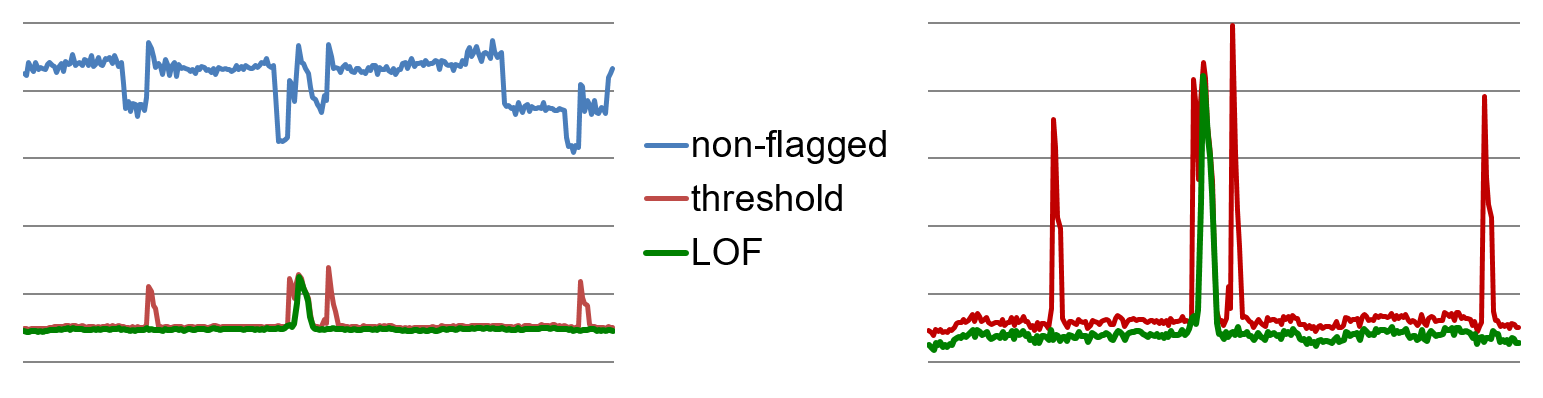}
\caption{Folded pulse profiles with and without flagging.}
\label{fig:profiles}
\end{figure}

In Figure~\ref{fig:profiles} (left side), we show the folded pulse
profiles, without RFI mitigation, with a simple thresholder, and with
LOF. Without RFI mitigation, the pulsar signal is completely below the
noise floor, and cannot be detected. With the thresholder, the pulse
is visible, but there also are false positives, caused by strong RFI
events that were not removed. With LOF, there is one clear peak with
the right shape, with a good SNR. This can be seen better in the right
side of Figure~\ref{fig:profiles}, which shows the same data, but
zoomed in, with the non-flagged line removed.

\begin{figure}[htb]
\center
\includegraphics[width=0.8\columnwidth]{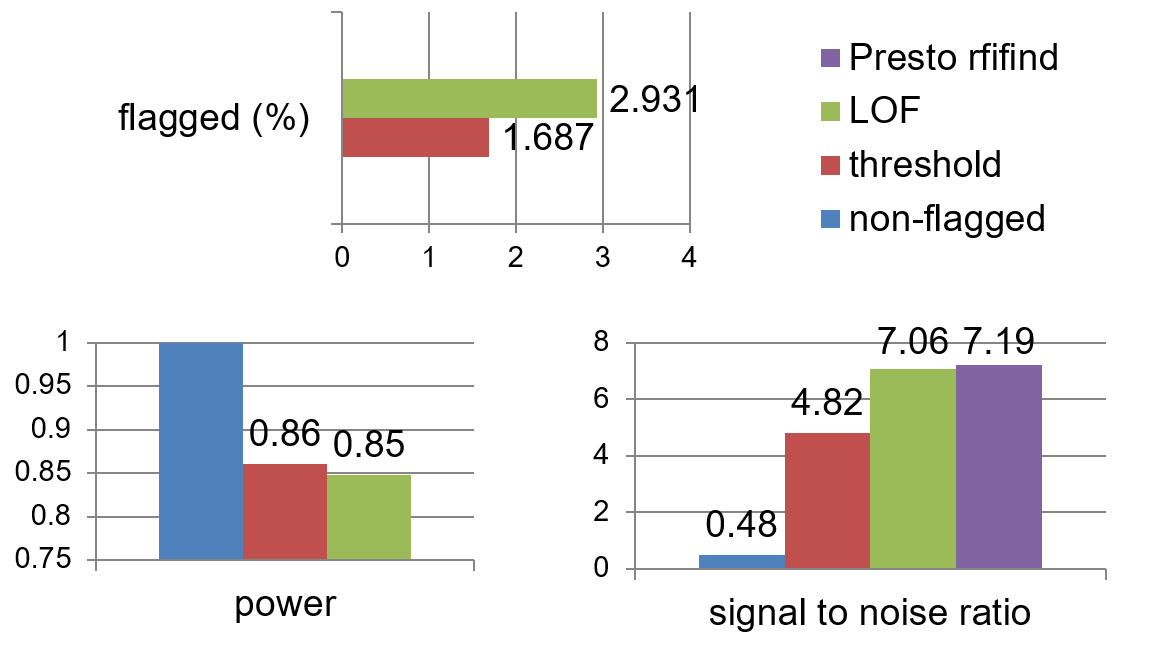}
\caption{Quality of LOF for pulsar B1919+21. LOF has significantly higher quality than simple thresholding. The quality of LOF is close to \emph{offline} Presto rfifind.}
\label{fig:snr}
\end{figure}

Figure~\ref{fig:snr} shows that LOF flags 2.9\% of the data in the observation,
significantly more than the simple thresholder that only flags
1.7\%. In the bottom left graph, we show that LOF flags away about
15\% of the total signal power, only slightly more than the simple
thresholder. The bottom right graph shows the SNR of the folded pulse
profile. LOF is significantly better than the simple thresholder, and
almost as high as performing \emph{offline} RFI mitigation with Presto's
rfifind~\cite{presto}.

\section {TOWARDS EXASCALE}

The LOFAR online processor needs hundreds of teraflops of computational
power. Future instruments such as the Square Kilometre Array (SKA)
will be much more sensitive, and will require orders of magnitude
more processing~\cite{ska}. Therefore, we were careful to make sure that all
algorithms used in our real-time flagger have
a linear computational complexity, allowing excellent scaling. In
addition, we investigated the use of modern processing architectures,
such as GPUs.  For LOFAR, we already switched from an IBM Blue Gene/p system to a GPU
cluster. For the central signal processor of the SKA, a combination of
FPGAs and GPUs are likely. In this section, we present a prototype GPU
version of the LOF.

\subsection {GPU implementation}
\label{sec:gpu}

\begin{figure}[htb]
\center
\includegraphics[width=0.8\columnwidth]{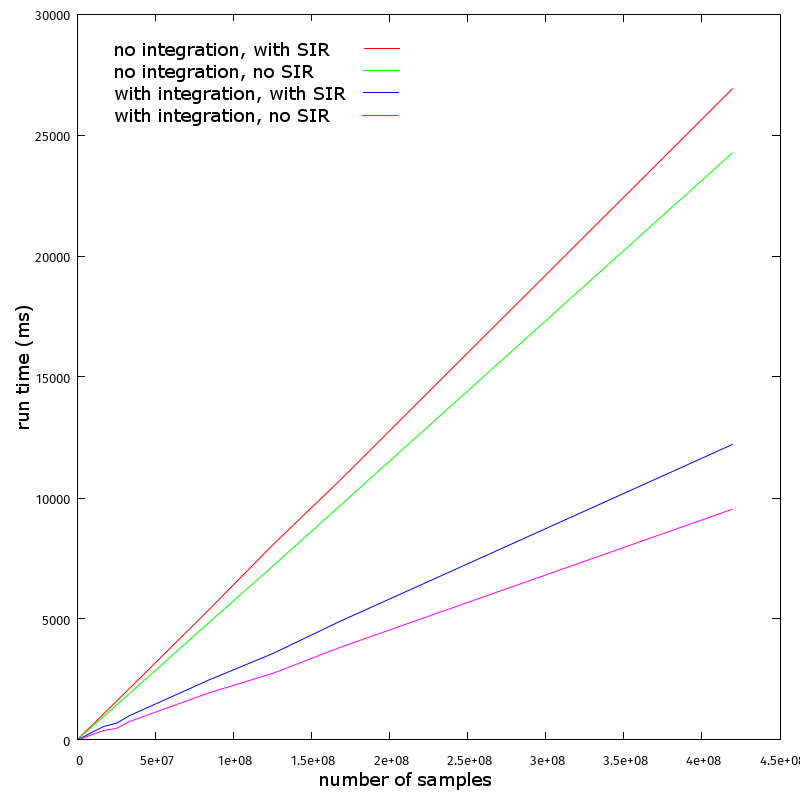}
\caption{Scalability of the GPU implementation of LOF.}
\label{fig:gpu}
\end{figure}

With Linus Schoemaker~\cite{thesis-linus}, we worked on the GPU port
of LOF. Due to the limited space, we will only describe the most
important differences with the normal parallel CPU version here.
The normal LOF exploits parallelism by scheduling different subbands to different compute
nodes. Inside the compute nodes we use C++ and OpenMP for
multi-threading, and asynchronous MPI messages for inter-node communication.
To avoid synchronization and parallelization
overhead, we make the parallelism as coarse grained as possible. This
means that each thread handles all data for one subband.

GPUs, in contrast, can efficiently exploit extremely fine grained
parallelism. Hundreds of thousands of threads can work in parallel on
a single GPU, without overhead.  In fact, in our implementation, we
create one GPU thread \emph{per sample}. We exploit data-reuse by
using the shared memory that is available inside the GPU's streaming
multiprocessors. The performance results are shown in
figure~\ref{fig:gpu}.  The top two lines (red and green) show the run
time without doing any data integration, running SumThreshold on the
full input data rate.  For the red line, we also run the SIR operator;
green is without.  The bottom two lines show performance if we fully
integrate and run SumThreshold two times, once in the frequency
direction, and once in the time direction, as described in
Section~\ref{sec:rt-changes}. In all cases, we achieve linear
Scalability. The GPU performs so well, that we can handle all LOFAR
stations on single GPU in real time.

\subsection {Integration in SKA time domain pipeline}

We are currently working on incorporating our CPU and GPU codes in the
SKA Central Signal Processor (CSP), in the context of the Time Domain
Team (TDT) pulsar and transient pipeline.  The initial stages of the
CSP, including the beam former, will likely use FPGAs. After the
beam former, GPUs are the most likely candidate for further
processing. Therefore, we will perform post-beamforming flagging using
our algorithms on GPUs. For an FPGA implementation, more research is needed.

\section{CONCLUSIONS AND FUTURE WORK}

We have demonstrated that our online flagger can achieve much higher
quality than simple thresholding, in real time, even on a distributed
system.  The SumThreshold algorithm was originally used mostly on
visibility data. In this paper, we demonstrated that the algorithm
also works well on raw voltages, pre-correlation data, and post-beam
forming data. Therefore, we have one robust algorithm for extremely
different scales, from microseconds to multiple seconds.  The
algorithms are scalable and have linear computational complexity,
adding little overhead to existing pipelines.  One complication is
that we have an extremely limited view on our data, and therefore need
a history flagger. Due to the high data rates, we have to be flexible
in storage requirements, even for statistics.

We are currently working on commissioning of the GPU code for
LOFAR. Moreover, we are constructing a performance model that can
extrapolate Scalability towards SKA sizes. This model includes 
power dissipation as well, since this will be an important bottleneck
for the SKA. We plan to use the Dome ExaBounds tool for this analysis~\cite{dome}.
All code used in this paper is available as open source: \\
\url{https://github.com/NLeSC/eAstroViz}.

\footnotesize{
\bibliographystyle{IEEEbib}
\bibliography{strings,refs}

\begin{thebibliography}{1}

\bibitem{andre-post}
A.~R. Offringa, A.~G. de~Bruyn, and S.~Zaroubi,
\newblock ``Post-correlation filtering techniques for off-axis source and rfi
  removal,''
\newblock {\em Monthly Notices of the Royal Astronomical Society}, vol. 422,
  no. 1, pp. 563--580, 2012.

\bibitem{vandeGronde2016}
Jasper~J. van~de Gronde, Andr{\'e}~R. Offringa, and Jos B. T.~M. Roerdink,
\newblock ``Efficient and robust path openings using the scale-invariant rank
  operator,''
\newblock {\em Journal of Math. Imaging and Vision}, vol. 56, no. 3, pp.
  455--471, 2016.

\bibitem{lofar-results}
A.R. Offringa, A.G. de~Bruyn, S.~Zaroubi, and M.~Biehl,
\newblock ``{A LOFAR RFI detection pipeline and its first results},''
\newblock in {\em Proc. of Science, RFI2010}, 2010, vol.~II, pp. 803--806.

\bibitem{lofar-performance}
John~W. Romein, P.~Chris Broekema, Jan~David Mol, and Rob~V. van Nieuwpoort,
\newblock ``{The LOFAR Correlator: Implementation and Performance Analysis},''
\newblock in {\em Proc. ACM Symposium on Principles and Practice of Parallel
  Programming (PPoPP'10)}, 2010, pp. 169--178.

\bibitem{presto}
S.~M. {Ransom}, S.~S. {Eikenberry}, and J.~{Middleditch},
\newblock ``{Fourier Techniques for Very Long Astrophysical Time-Series
  Analysis},''
\newblock {\em The Astronomical Journal}, vol. 124, pp. 1788--1809, Sept. 2002.

\bibitem{ska}
P.~Chris Broekema, Rob~V. van Nieuwpoort, and Henri~E. Bal,
\newblock ``The square kilometre array science data processor preliminary
  compute platform design,''
\newblock {\em Journal of Instrumentation}, July 2015.

\bibitem{thesis-linus}
Linus Schoemaker,
\newblock ``{Removing Radio Frequency Interference in the LOFAR using GPUs},''
\newblock M.S. thesis, VU University Amsterdam, 2015.

\bibitem{dome}
P.~Chris Broekema, Albert-Jan Boonstra, Victoria~Caparrós Cabezas, Ton
  Engbersen, Robert Haas, Hanno Holties, Jens Jelitto, Ronald~P. Luijten, Peter
  Maat, Rob~V. van Nieuwpoort, Ronald Nijboer, John~W. Romein, and Bert~Jan
  Offrein,
\newblock ``{DOME: Towards the ASTRON \& IBM Center for ExaScale Technology},''
\newblock in {\em Proc. ACM workshop on AstroHPC}, 2012.

\end{thebibliography}
}
\end{document}